\documentclass{elsart}

\usepackage{amssymb}
\usepackage{graphicx}
\begin{document}

\begin{frontmatter}



\title{Fake one-time pad cannot be used to improve the efficiency of quantum communication}


\author[BUPT,XD]{Fei Gao\corauthref{cor}},
\corauth[cor]{Corresponding author.} \ead{hzpe@sohu.com}
\author[BUPT]{Su-Juan Qin},
\author[BUPT]{Qiao-Yan Wen},
\author[CD]{Fu-Chen Zhu}

\address[BUPT]{School of Science, Beijing University of Posts and Telecommunications, Beijing, 100876, China}
\address[XD]{State Key Laboratory of Integrated Services Network, Xidian University, Xi'an, 710071, China}
\address[CD]{National Laboratory for Modern Communications, P.O.Box 810, Chengdu, 610041, China}

\begin{abstract}
Two misuses of one-time pad in improving the efficiency of quantum
communication are pointed out. One happens when using some message
bits to encrypt others, the other exists because the key bits are
not truly random. Both of them result in the decrease of security.
Therefore, one-time pad should be used carefully in designing
quantum communication protocols.
\end{abstract}

\begin{keyword}
quantum key distribution \sep quantum secure direct communication
\sep quantum cryptography \sep one-time pad

\PACS 03.67.Dd \sep 03.67.Hk \sep 03.65.-w
\end{keyword}
\end{frontmatter}

The aim of cryptography is to ensure that a secret message is
transmitted between two users in a way that any eavesdropper
cannot read it. In classical cryptography, it is generally
accepted that one-time pad \cite{Vernam}, which utilizes a
previously shared secret key to encrypt the message transmitted in
the public channel, is the only proved secure cryptosystem
\cite{Shannon}. However, it is difficult for all existing
classical cryptosystems to establish a random key with
unconditional security between the users. Fortunately, quantum key
distribution (QKD) \cite{BB84,E91,B92,GRTZ02}, the approach using
quantum mechanics principles for the distribution of secret key,
can overcome this obstacle skillfully. Since both QKD and one-time
pad have proved security \cite{L96,LC99,SP00,GL03}, the
cryptosystem of ``QKD \& one-time pad'' is a perfect one when the
security is concerned.

Quantum secure direct communication (QSDC)
\cite{BF02,DLL03,DL04,LM05} is another branch of quantum
cryptography. Different from QKD, QSDC allows the sender transmits
directly the secret (not a random key) to the receiver in a
deterministic and secure manner. If it is designed carefully, a
QSDC protocol can also attain unconditional security \cite{DLL05}.

With the fast development of quantum cryptography, more and more
novel QKD and QSDC protocols were proposed. An important criterion
of a protocol is its efficiency. In the work of scheme designing,
a higher efficiency is the goal the scheme-designers always
pursue. However, the feasibility of some ways that lead to high
efficiency should be reexamined. In this Letter we choose two
typical protocols to discuss, where the alleged high efficiency is
illusory. That is, such an unrealistic efficiency would result in
insecurity.

Recently, a semi-direct quantum secure communication protocol was
presented in Ref. \cite{An07}. In this protocol, three users
Alice, Bob, and Charlie can exchange one bit of message securely
by using one GHZ-state. That is, each of them can send one bit of
message to the other two person while the outside eavesdropper,
say Eve, can never obtain any information about these bits.
Setting aside the particular process of this part of protocol, we
only discuss its way used to improve the efficiency. In the
following discussion, we assume that the bits transmitted by
quantum process are unconditionally secure. As we can see, the
efficiency of the quantum process is one bit per GHZ-state. To
make it more efficient, the users employ the quantum process to
send the odd-numbered secret bits but a classical one to send the
even-numbered bits, namely, using the odd-numbered bits to encrypt
the even-numbered ones. For example, suppose Alice's secret bit
string is $\{a_1,a_2,a_3,...,a_N\}$. Alice first sends $a_1$ to
Bob and Charlie by quantum process. Afterwards, Alice calculates
$a^{'}_{1}=a_1\oplus a_2$ ($\oplus$ denotes the addition modulo 2)
and publicly broadcasts $a^{'}_{1}$. With the knowledge of $a_1$,
Bob and Charlie can deduce $a_2$ just by $a_2=a^{'}_{1}\oplus
a_1$. On the contrary, as an outside eavesdropper, Eve cannot
obtain any information about $a_2$ since she does not know $a_1$.
Similarly, Alice sends $a_3$ by quantum process and $a_4$ by
classical one, and so on. As the author of Ref. \cite{An07}
pointed out, each of the odd-numbered secret bits ``is used only
once in the encoding so their confidentiality has not been leaked
out at all, in accordance with the one-time pad.'' By this way,
the efficiency of whole protocol is doubled to 2 bits per
GHZ-state.

Indeed, the above process to send the even-numbered secret bits
looks like that of one-time pad. However, it is really not so.
Consider the scenario where Alice and Bob share two secure key
bits $\{k_1,k_2\}$ and Alice wants to send two confidential bits
$\{p_1,p_2\}$ to Bob. In a real one-time pad, Alice encrypts the
plaintext $\{p_1,p_2\}$ with the key bits $\{k_1,k_2\}$, obtaining
the ciphertext $\{c_1,c_2\}=\{p_1\oplus k_1,p_2\oplus k_2\}$.
Afterwards, Alice sends $\{c_1,c_2\}$ to Bob publicly. With the
knowledge of $\{k_1,k_2\}$, Bob can obtain the plaintext by the
decryption $\{p_1,p_2\}=\{c_1\oplus k_1,c_2\oplus k_2\}$. On the
contrary, Eve cannot extract any information about the plaintext
from $\{c_1,c_2\}$. As a result, Alice can transmit 2 bits to Bob
securely by the above process. Differently, in Ref. \cite{An07},
Alice uses the first \emph{message bit} $a_1$ instead of a key bit
to encrypt the second one $a_2$ and broadcasts the ciphertext
$a^{'}_{1}$, which results in an invalid transmission from the
perspective of information theory and cryptography. That is, Eve
can attain some information about the message bits $\{a_1,a_2\}$
from the declared $a^{'}_{1}$. For example, $a^{'}_{1}=0$. Then
Eve knows either $\{a_1,a_2\}=00$ or $\{a_1,a_2\}=11$, which
contains only $\log_22=1$ bit of information for her. When
$a^{'}_{1}=1$ we will draw the similar conclusion. Therefore, from
the declared $a^{'}_{1}$ Eve can obtain 1 bit of information about
the two message bits though she does not know the particular value
of them. Consequently, every time Alice sends two message bits
according to the protocol in Ref. \cite{An07}, Bob obtains all the
2 bits of information while Eve can also get 1. Namely, in
essence, Alice just transmits 1 bit of information to Bob
\emph{securely}. From this point of view, the way to improve
efficiency in Ref. \cite{An07} is null.

Things do not come singly but in pairs. Not long ago, Li
\textit{et al}. presented a QKD scheme \cite{CL03} based on
entanglement swapping \cite{ES}. In this protocol, Alice and Bob
previously shared enough EPR pairs in known states. Without loss
of generality, consider two pairs
$|\Phi^+\rangle_{AB}^{12}=1/\sqrt{2}(|00\rangle+|11\rangle)$ and
$|\Psi^+\rangle_{AB}^{34}=1/\sqrt{2}(|01\rangle+|10\rangle)$,
where the superscripts 1, 2, 3, 4 denote the different particles.
Alice holds particles 1, 3 and Bob controls 2, 4. When they
distribute key bits, Alice and Bob perform entanglement swapping
between these two EPR pairs. According to the rule of entanglement
swapping, each of them knows not only his/her measurement result
but also his/her counterpart's. The author of Ref. \cite{CL03}
alleged that these two results can bring four key bits to Alice
and Bob. For example, when Alice performs a Bell measurement on
particles 1 and 3 she gets $|\Psi^+\rangle_{AA}^{13}$, she can
deduce that Bob's measurement outcome must be
$|\Phi^+\rangle_{BB}^{24}$. If four EPR states $|\Phi^+\rangle$,
$|\Phi^-\rangle$, $|\Psi^+\rangle$, and $|\Psi^-\rangle$ represent
00, 01, 10, and 11 respectively, Alice will obtain four key bits
1000, where 10 comes from $|\Psi^+\rangle_{AA}^{13}$ and 00
corresponds to $|\Phi^+\rangle_{BB}^{24}$. At the same time, Bob
can attain these four key bits by similar deduction. Therefore,
four particles bring four key bits, which means a double
efficiency of that in BB84 protocol \cite{BB84}.

However, the efficiency may not be so high. As we know, for two
given EPR pairs, the two measurement results in entanglement
swapping are not completely random. On the contrary, they have
strong correlation. Consider the above example again, because
\begin{eqnarray}
|\Phi^+\rangle_{AB}^{12}|\Psi^+\rangle_{AB}^{34}=\frac{1}{2}\{|\Phi^+\rangle_{AB}^{13}|\Psi^+\rangle_{AB}^{24}+|\Phi^-\rangle_{AB}^{13}|\Psi^-\rangle_{AB}^{24}\nonumber\\
+|\Psi^+\rangle_{AB}^{13}|\Phi^+\rangle_{AB}^{24}+|\Psi^-\rangle_{AB}^{13}|\Phi^-\rangle_{AB}^{24}\}
\end{eqnarray}

anyone, including Eve, who knows the initial state
$|\Phi^+\rangle_{AB}^{12}|\Psi^+\rangle_{AB}^{34}$ can draw a
conclusion that Alice and Bob's outcome pair must be one of
$\{|\Phi^+\rangle_{AB}^{13}|\Psi^+\rangle_{AB}^{24}$,
$|\Phi^-\rangle_{AB}^{13}|\Psi^-\rangle_{AB}^{24}$,
$|\Psi^+\rangle_{AB}^{13}|\Phi^+\rangle_{AB}^{24}$,
$|\Psi^-\rangle_{AB}^{13}|\Phi^-\rangle_{AB}^{24}\}$ randomly.
Therefore, Eve knows the key bits Alice and Bob obtain should be
one of \{0010, 0111, 1000, 1101\} with equal probability while
other twelve results (such as 0000, 0100, etc) never appear, which
contains \emph{only} $I=-\sum_{i=1}^4\frac{1}{4}\log\frac{1}{4}=2$
bits of information. As a result, if Alice encrypts her secret by
using above results as four key bits of one-time pad, it would be
leaked partly to Eve when the ciphertext is transmitted publicly.
For example, let $\{p_1, p_2, p_3, p_4\}$ and $\{k_1, k_2, k_3,
k_4\}$ denote four bits of plaintext and key respectively. Then
the ciphertext equals $\{c_1, c_2, c_3, c_4\}=\{p_1\oplus k_1,
p_2\oplus k_2, p_3\oplus k_3, p_4\oplus k_4\}$. Observing the
possible values of the above key bits, we can see that $k_1\oplus
k_3=1$ and $k_2\oplus k_4=0$ always hold. Sequently, when the
cipertext is announced, Eve always know that $p_1\oplus
p_3=c_1\oplus k_1\oplus c_3\oplus k_3=c_1\oplus c_3\oplus 1$, and
$p_2\oplus p_4=c_2\oplus k_2\oplus c_4\oplus k_4=c_2\oplus c_4$,
which implies a two-bit leakage of the secret. In a word,  to
attain security, the efficiency of the protocol in Ref.
\cite{CL03} should be 2 bits per entanglement swapping, but not
the alleged 4.

Both the above errors are related to the right understanding of
one-time pad. It was shown, by Shannon \cite{Shannon}, that the
one-time pad which meets the following three conditions is
perfectly secure: (i) the key is truly random, (ii) the key has
the same length as the message, (iii) the key is never reused. In
Ref. \cite{An07}, the user use a \emph{message} bit, but not a key
bit, to encrypt another one. In Ref. \cite{CL03}, the key bits are
\emph{correlated} but not truly random. Neither of them is a real
one-time pad. Therefore, we should know not only one-time pad can
achieve perfect security but also the requirements to possess this
merit. We emphasize that unconditional security is a crucial
feature of quantum cryptography (generally QKD \& one-time pad)
and it should never be sacrificed to improve the performance of
other aspects such as efficiency.

As we analyzed above, fake one-time pad cannot be used to improve
the efficiency of a quantum communication protocol. In fact the
efficiency was bounded by Holeve quantity \cite{QCQI}, which
implies that $n$ qubits cannot be used to transmit more than $n$
bits of classical information. So, 1 key bit per qubit is already
the full efficiency. In a 2-level system it equals 1 bit per
particle (here we do not discuss $d$-level quantum system
\cite{KBB02,GGW05,GGW06}, which can certainly reach higher
efficiency than a 2-level one). For example, with the qubit
storage facility, the delayed-choice BB84 protocol \cite{GGWZ06}
can achieve full efficiency in theory. From this point of view,
the alleged high efficiencies in both Ref. \cite{An07} and
\cite{CL03} are illusory because they are even exceed a maximal
value which is allowed by quantum mechanics.

In summary, we point out two misuses of one-time pad in improving
the efficiency of quantum communication \cite{An07,CL03}. Indeed,
one-time pad can accomplish perfect security. But we should always
remember its necessary conditions when we utilize it in quantum
cryptography. Otherwise, the quantum protocol may become insecure.

This work was supported by the National Natural Science Foundation
of China, Grant No. 60373059; the National Laboratory for Modern
Communications Science Foundation of China, Grants No.
9140C1101010601; the National Research Foundation for the Doctoral
Program of Higher Education of China, Grant No.20040013007; the
Major Research plan of the National Natural Science Foundation of
China, Grant No. 90604023; the Graduate Students Innovation
Foundation of BUPT; and the ISN Open Foundation.

\end{document}